\documentclass[conference]{IEEEtran}
\IEEEoverridecommandlockouts
\usepackage[whole]{bxcjkjatype}
\usepackage{cite}
\usepackage{amsmath,amssymb,amsfonts}
\usepackage{algorithmic}
\usepackage{graphicx}
\usepackage{textcomp}
\usepackage{xcolor}
\usepackage{multirow}
\usepackage{comment}
\usepackage{url}
\usepackage{breqn}

\usepackage[caption=false]{subfig}

\def\BibTeX{{\rm B\kern-.05em{\sc i\kern-.025em b}\kern-.08em
    T\kern-.1667em\lower.7ex\hbox{E}\kern-.125emX}}

\begin{document}

\title{The metrics of keywords \\to understand the difference \\between Retweet and Like in each category}
\author{
    \IEEEauthorblockN{Kenshin Sekimoto\IEEEauthorrefmark{1} \quad Yoshifumi Seki\IEEEauthorrefmark{2} \quad Mitsuo Yoshida\IEEEauthorrefmark{1} \quad Kyoji Umemura\IEEEauthorrefmark{1}}
    \IEEEauthorblockA{\IEEEauthorrefmark{1}\textit{Toyohashi University of Technology}\\
    Aichi, Japan\\
    sekimoto.kenshin.lo@tut.jp, yoshida@cs.tut.ac.jp, umemura@tut.jp}
    \IEEEauthorblockA{\IEEEauthorrefmark{2}\textit{Gunosy Inc.}\\
    Tokyo, Japan\\
    yoshifumi.seki@gunosy.com}
}

\maketitle

\begin{abstract}
The purpose of this study is to clarify what kind of news is easily retweeted and what kind of news is easily Liked.
We believe these actions, retweeting and Liking, have different meanings for users.
Understanding this difference is important for understanding people's interest in Twitter.
To analyze the difference between retweets (RT) and Likes on Twitter in detail, we focus on word appearances in news titles.
First, we calculate basic statistics and confirm that tweets containing news URLs have different RT and Like tendencies compared to other tweets.
Next, we compared RTs and Likes for each category and confirmed that the tendency of categories is different.
Therefore, we propose metrics for clarifying the differences in each action for each category used in the $\chi$-square test in order to perform an analysis focusing on the topic.
The proposed metrics are more useful than simple counts and TF--IDF for extracting meaningful words to understand the difference between RTs and Likes.
We analyzed each category using the proposed metrics and quantitatively confirmed that the difference in the role of retweeting and Liking appeared in the content depending on the category.
Moreover, by aggregating tweets chronologically, the results showed the trend of RT and Like as a list of words and clarified how the characteristic words of each week were related to current events for retweeting and Liking.
\end{abstract}

\section{Introduction}

Analyzing social media, such as Twitter, Facebook, and Instagram, is an important method for understanding what people are interested in.
Social media services have many functions, and analyzing functions is an important way to glean knowledge about user interests and social trends.
In particular, because most of these social media have a favorite function and a sharing function for posts, they can be analyzed effectively.
Therefore, in this paper, we target Twitter, which is representative of social media. 
A retweet (RT) is a function that allows users to share news, tweets, and other information that was already shared by others, whereas a Like is a function that allows users to share their positive feelings about a certain tweet.
We believe these actions, retweeting and Liking, have different meanings for users.
Understanding this difference is important for understanding people's interest in social media, especially Twitter.

In this study, to understand people's interest in using Twitter, we attempt to clarify what kind of news is easily retweeted and what kind of news is easily Liked.
Social media content is generally noisy because there are many unique expressions by users, and it is difficult to understand the content.
Conversely, news content is relatively easy to understand because it is properly organized, with easy-to-understand titles and uniformed categories.
A lot of news media try to utilize social media to enhance media influence and to improve their profits.
Therefore, it is important to analyze social media content, including news content.
Since people engage in retweeting and Liking based on their interests, we assume that people's interests can be understood by analyzing news spreading on social media platforms.
In this paper, since large data can be used for the analysis using the Search API, we use Twitter data. 
Thus, this study focuses on clarifying the difference between RT and Like using news sharing on Twitter.

To analyze the difference between RTs and Likes on Twitter in more detail, we focus on word appearances in news titles.
The words included in the title are the simplest metrics for substituting topics.
If the words included in the title are different between the retweeted and Liked tweets, it is suggested that the meaning of each action differs for users.

We propose a metric using the expected frequency used in the $\chi$-square test in order to perform an analysis focusing on this topic.
The simplest method for obtaining characteristic words is by counting the frequency of word usage. 
However, using this method, it is difficult to obtain effective knowledge because many common words are extracted.
Although TF--IDF is a well-known method for solving this problem, it is difficult to use because tweets texts are short and the number of tweets is huge.
In this paper, we confirm quantitatively whether the proposed metrics are better than simple counting and TF--IDF.

First, we calculated basic statistics, and confirmed that tweets containing news URLs have different RT and Like tendencies compared to other tweets.
It was suggested that the user might behave differently when faced with tweets including news URLs (NewsURLs).
Next, we compared RTs and Likes for each category and confirmed that the tendency of categories is different.
Finally, we proposed metrics and analyzed one year of news and tweet data using proposed-metrics.
From the results of the analysis, we clarified the following points:

\begin{itemize}
  \item With the proposed metrics, meaningless words were removed from the results of the extract, and words related to categories could be extracted.
  \item When looking at the extracted words, many categories could be partially interpreted, confirming the difference between RTs and Likes.
  \item Using metrics to extract words from tweets by chronological order, we can obtain and interpret the trend of RT and Like as a list of words.
\end{itemize}

The achievements of this paper are as follows:

\begin{itemize}
    \item We compared RTs and Likes for each category using huge news tweet datasets and confirmed that the tendency of categories is different.
    \item We proposed metrics for clarifying the difference in social media actions for each category.
    \item We analyzed each category using the proposed metrics and quantitatively confirmed that the difference in the role of retweeting and Liking appeared in the content depending on the category.
    \item The results indicated that it is possible to obtain keywords for grasping past trend tendencies by aggregating by time series and using proposed metrics, depending on the category.
\end{itemize}

\section{Related Work}

Analyzing social media's user action for posts, such as comment, Like, and so on, is one of the topics and research is active.
For example, in the studies about social media's search system by using a search model based on information on user's actions and polarity of comment, they succeed to improve than baseline~\cite{badache2019, badache2015}. 
The purpose of this paper is to clarify the difference between tweets that are easily retweeted and tweets that are easily Liked in the news domain.
First, we describe the studies on RT and Like.
After that, we describe the studies dealing with both RT and Like.
Finally, we introduce studies on news in social media.

Retweeting is an active research topic in studies on Twitter~\cite{Mohammadi2018}.
A study about retweeting analyzed user behavior related to retweeting\cite{Xu2012}, analyzed information diffusion, predicted how much a tweet would be retweeted, and so on.
A study~\cite{Suh2010} on the ease of retweeting revealed that tweets including  ``hash tags'' and URLs are more diffused than other tweets and the number of tweets of the user have little impact on RT tendency.

Liking is a feature formerly called favoriting, and though it is not as common as  RTs, there are some studies about Likes.
One study analyzed user behavior with regard to favorite features, comparing Twitter with Fickler~\cite{Lee2010}.
Another study applied the method of network analysis to the Favorite, Follow, and Mention functions on Twitter~\cite{Kato2012}.
A large-scale survey sought to understand the motivation of using the Favorite function on Twitter~\cite{Meier2014}.

There are various studies about user motivation related to both retweeting and Liking, but there are few studies on the difference between retweeting and Liking regarding tweeted content.
Previously, Liking was a user's personal favorite feature and did not affect the spread of tweets, but recently, liked tweets have been displayed to other users, so Liking has become a more important action.
Understanding the characteristics of retweeting and Liking may become important in the future.
Moreover, 
Recently, the relationship between news and social media has become important, and there are many studies about this relationship.
Two studies~\cite{Aldous2019a,Aldous2019b} focused on the number of comments and the number of favorites for news on social media and found that the news content (e.g. image, video, and body) affects sharing behavior across multiple social media platforms.
Another study~\cite{Trilling2017} attempted to predict the sharing of news over social media using the features of news articles.
Further research~\cite{Fletcher2017} analyzed the impact of trust in news media on online news consumption, viewing, commenting, and sharing on social media.
Thus, the relationship between news and social media is an active research topic, and we assume that our study is also important because of the metrics proposed in this paper help these analyses.

\section{Dataset}
\subsection{News data}
The news data used in this study included 355086 articles collected from the news portal site Ceek.jp News\footnote{\url{http://news.ceek.jp/} (accessed 2019--12--24)} during the 12 months from January 1, 2017, to December 31, 2017.
All of the news data were written entirely in Japanese.
In this study, we used the news URL, news title, and news category for each news article.
Each item belonged to only one category, and there were 12 news categories: economy, entertainment, etc., IT, local, society, politics, science, sports, world, obituaries, and CN--KR (China and Korea).

\subsection{Tweet data}
\begin{figure}[tb]
    \centering
    \includegraphics[scale=0.5]{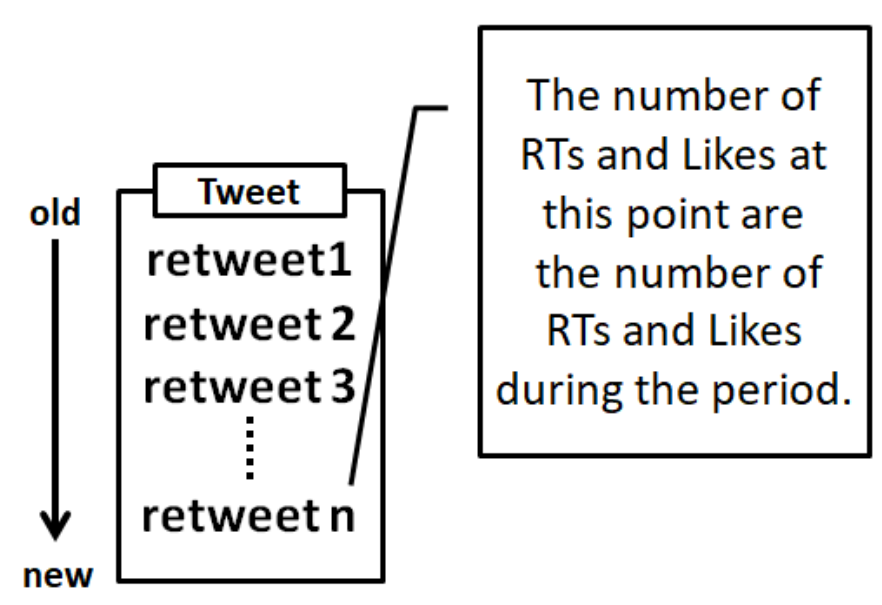}
    \caption{\textbf{How to count the number of RTs and Likes:} the RT count and Like count at the time of the last retweet in the aggregation period are the latest RT and Like counts in the dataset period.}
    \label{fig:countrtfav}
\end{figure}

All of the tweet data used in this study were Japanese retweet data collected from January 1, 2017, to December 31, 2017, using Twitter Search API\footnote{\url{https://developer.twitter.com/ja/docs/ads/general/api-reference}\\(accessed 2019--12--24)}.
Each tweet had various attribute values\footnote{\url{https://developer.twitter.com/en/docs/tweets/data-dictionary/overview/tweet-object} (accessed 2019--12--24)}, among which the tweet ID ($ id $), the user name ($ screen\_name $), the tweet date and time ($ created\_at $), the number of RTs ($ retweet\_count $), the number of Likes ($ favorite\_count $), the body of the tweet ($ text $), and the URL in the tweet ($ expand\_url $) were extracted and used for the analysis.
As shown in Fig.~\ref{fig:countrtfav}, the RT counts and Like counts at the time of the last retweet in the aggregation period are the latest retweet and Like counts in the tweet period.
Even if a RT was deleted, our count data cannot reflect this elimination.     

\subsection{Extraction of mentioned tweets}
To analyze tweets including news URLs, we filtered tweets using news data.
We collected the tweets with URLs from the tweet dataset and extracted the tweets with URLs included in the news dataset.
One tweet may refer to multiple news articles, and in this case, the tweet was associated with each news article.

\begin{figure}[tb]
    \centering
    \includegraphics[scale=0.15]{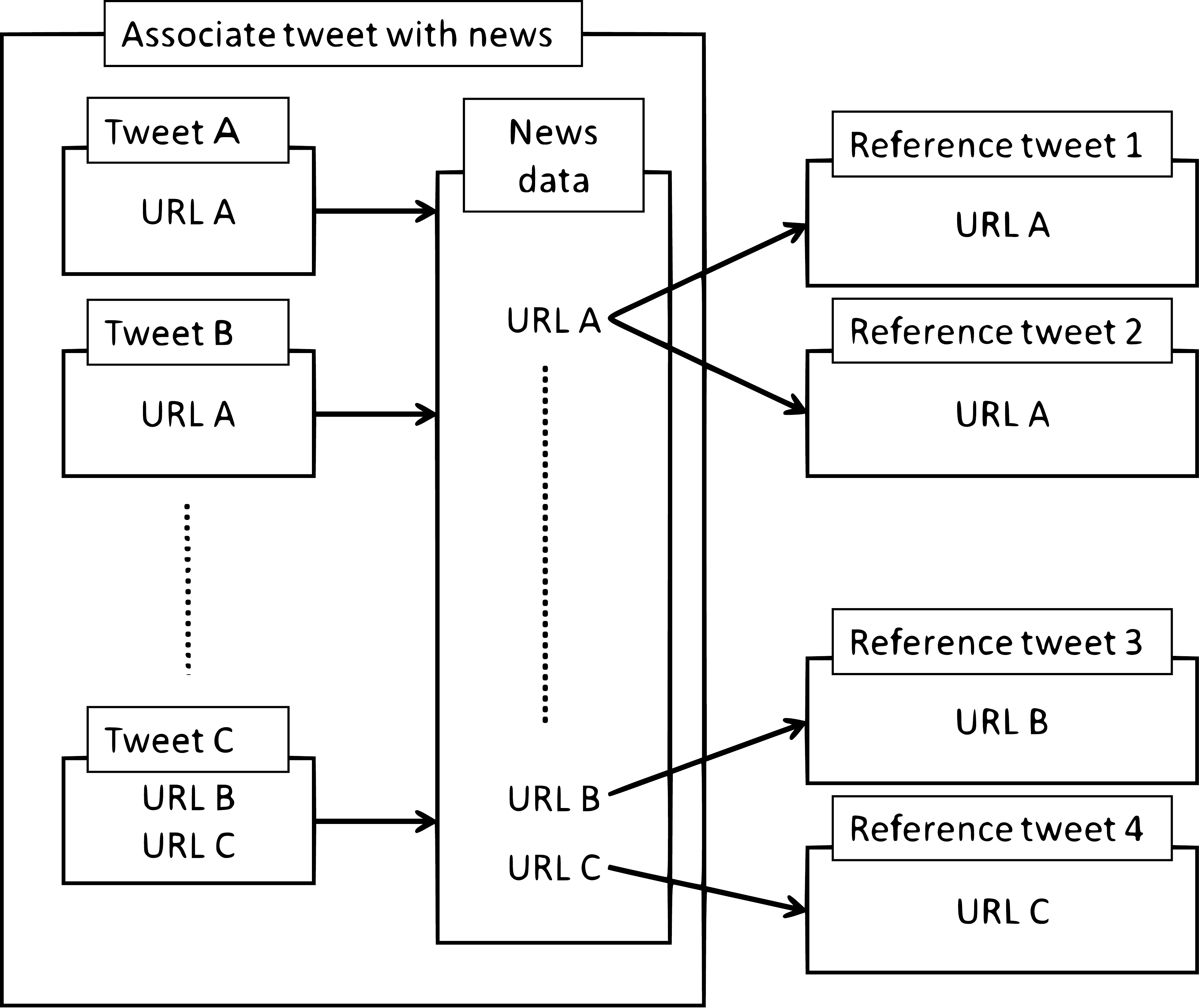}
    \caption{\textbf{Example of reference tweets:} The tweets have URLs included in the news dataset. A single tweet may refer to multiple news articles.}
    \label{fig:exdata}
\end{figure}

Fig.~\ref{fig:exdata} shows an example.
As a result of the matching procedure, the data matched 2615563 cases.
Because a single tweet may refer to multiple news articles, there were duplicates in the tweets of the corresponding data.

To exclude data that were rarely shared, such as tweets with 1 RT and 0 Likes, we limited the number of both RTs and Likes to 10 or more.
As a result, we obtained 567484 pairs of tweets and news.
In this paper, the 567384 pairs are referred to as ``reference tweets.''
Table~\ref{count of reference tweet} shows the number of data for each category of reference tweets.

In this study, the obituaries category has fewer tweets, and the CN--KR category encompasses many sensitive topics, so it is excluded from the analysis results.

\begin{table}[tb]
\caption{\textbf{The number of reference tweets in each category:} since the number of reference tweets in the entertainment category is the highest, it is often illustrated in this paper.} 
\centering
\begin{tabular}{lr}
Category & \# of reference tweets\\
\hline \hline
economy & 21462 \\
entertainment & 116600 \\
etc. & 68919 \\
IT & 104180 \\
local & 42749 \\
society & 65989 \\
politics & 53705 \\
science & 7986 \\
sports & 28681 \\
world & 53652 \\
\hline
CN--KR & 3203 \\
obituaries & 258 \\
\hline
\hline
Total & 567384 \\
\hline
\end{tabular}

\label{count of reference tweet}
\end{table}

\section{Preliminary Analysis}

\subsection{Basic statistics}

First, we compared tweets that include news URLs with all the tweets and the tweets that include URLs.
This comparison is to clarify the characteristics of tweets that include news URLs.
To identify the tweets including URLs, we utilized the URL entities of the Tweet Object from the Twitter API response.
However, when a user attaches images or videos to a tweet, these contents are uploaded on the Twitter server, and the tweet includes the Twitter domain URL in the URL entities.  
Thus, we filtered the tweets including the Twitter domain URL from the tweets that have the URL entities.

This comparison used tweets from January 1, 2017, to January 7, 2017, in our tweet dataset.
We call all the tweets {\it ALL}, the tweets including URLs {\it AllURLs}, and the tweets including news URL {\it NewsURLs} (\textit{NewsURL} $\subseteq$ \textit{AllURLs} $\subseteq$ \textit{ALL}).
For liked reference tweets, the number of both RTs and Likes was limited to 10 or more.

Table~\ref{statistic} shows the statistics of each tweet type.
The total number of ALL tweets was 2121431, the number of AllURL tweets was 567384, and the number of NewsURL tweets was 7367.
Compared with the averages of RTs and Likes of ALL and AllURLs, the averages of RTs are twice as large as those of Likes. 
On the other hand, in the case of NewsURLs, the averages of RTs and Likes were about the same.
Furthermore, this magnitude relationship is almost the same for any quartile.
From this result, NewsURL characteristics are differed from those of ALL or AllURL.

\begin{table}[tb]
  \caption{\textbf{The statistics of each tweet type:} the tweets including news URL (NewsURLs) characteristics are differed from those of the ALL or AllURLs.}
  \centering
  \resizebox{\columnwidth}{!}{
  \begin{tabular}{lrrrrrr}
        & \multicolumn{2}{c}{ALL} &  \multicolumn{2}{c}{AllURLs} & \multicolumn{2}{c}{NewsURLs} \\ \cline{2-7}
        &\multicolumn{1}{c}{RTs}  & \multicolumn{1}{c}{Likes} & \multicolumn{1}{c}{RTs} & \multicolumn{1}{c}{Likes} & \multicolumn{1}{c}{RTs} & \multicolumn{1}{c}{Likes} \\
  \hline
  mean  & 464.1 & 825.8 & 263.6 & 479.8 & 110.8 & 129.2 \\
  std   & 2922.1 & 4345.7 & 1870.6  & 2961.5 & 436.1 & 550.7 \\
  25\%  & 19 & 38 & 19 & 29 & 21 & 20 \\
  50\%  & 44 & 97 & 39 & 68 & 39 & 37 \\
  75\%  & 148 & 331 & 112  & 200 & 85 & 86 \\
  max   & 291159 & 502836 & 288029 & 455542 & 19020 & 21819 \\
  \hline
  \end{tabular}
  }
  \label{statistic}
\end{table}

\begin{figure}[tb]
    \centering
    \includegraphics[scale=0.43]{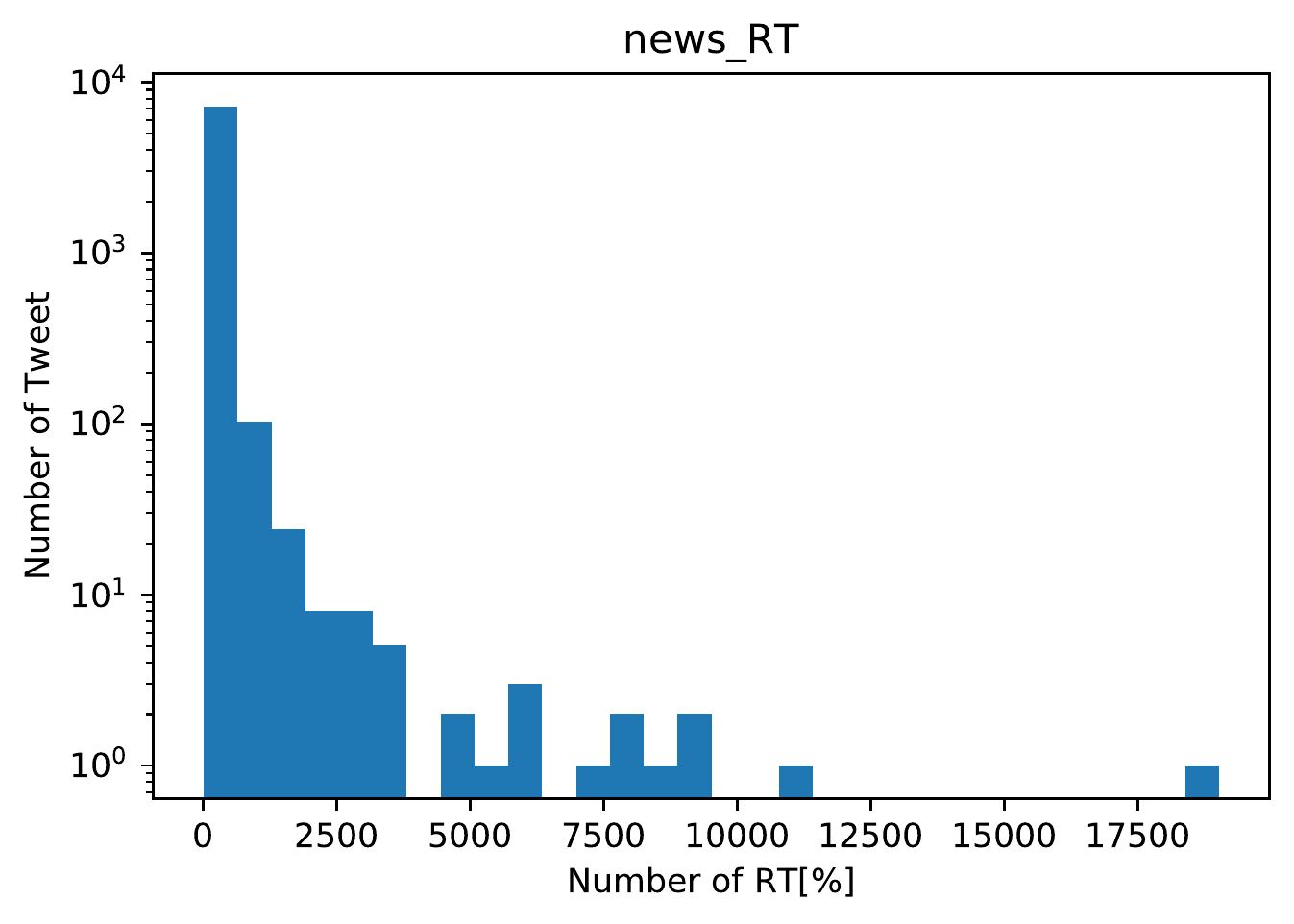}
    \includegraphics[scale=0.43]{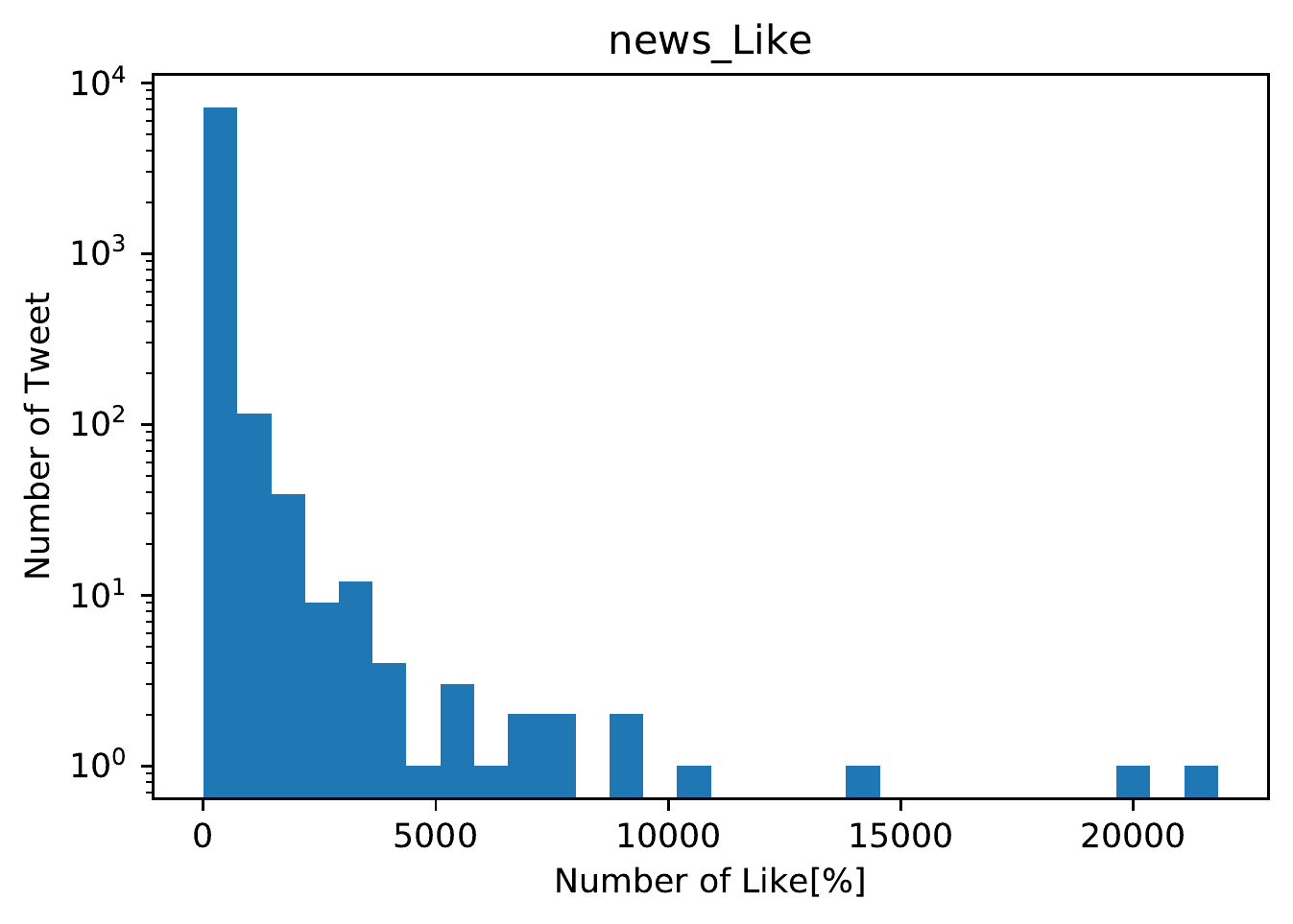}
    \caption{\textbf{Distribution of the number of RTs and Likes in NewsURLs:} General power law appeared in Like and RT, and this tendency was the same for ALL and AllURLs.}
    \label{distribution}
\end{figure}

Next, in Fig.~\ref{distribution}, we show the distribution of the number of RTs and Likes including NewsURLs.
The horizontal axis represents the distribution area of the number of RTs (news\_RT) and Likes (news\_Like), and the vertical axis represents the number of tweets plotted by Log-scale.
General power law appears in Likes and RTs, and this tendency was the same for ALL and AllURLs.

\subsection{Difference in retweeting and Liking in each category}

Next, we analyzed tweets, including news URLs, by news category.
In previous studies by focusing on the dwell time on news pages, it is clear that there is a difference in news browsing behavior for each category \cite{Seki2018}.
Based on this knowledge, we hypothesized that there was a difference between retweeting and Liking on Twitter for each category.
Thus, we introduced the metrics $RT\_Like$ to measure the difference in the ease of retweeting or Liking a set of tweets, such as news category.
$RT\_Like$ is defined as the ratio between the median RT ($median\_RT$) and the median Like ($median\_Like$). 
$RT\_Like$ indicates that is easy to be Liked if $RT\_Like$ it is smaller than 1.0, and it is easy to be retweeted if $RT\_Like$ is larger than 1.0:

\begin{gather}
    \mathrm{RT \_ Like} = \frac{ median\_RT }{ median\_Like }. \label{eq:rt_fav}
\end{gather}

Table~\ref{rt_fav} shows the results of the measurement of $RT\_Like$ for each category, and the categories that are likely to be retweeted are society, politics, and world.
The categories that are easily Liked are ``entertainment'' and ``IT.''
Therefore, it was confirmed that there was a difference between retweeting and Liking in each category.

  \begin{table}[tb]
  \caption{\textbf{Comparison between RTs and Likes in each category:} There was a difference between retweeting and Liking in each category.}
  \label{rt_fav}
  \centering
  \resizebox{\columnwidth}{!}{
  \begin{tabular}{lrrrr}
  Category & \# of Tweets & Median\_RT & Median\_Like & RT\_Like \\
  \hline \hline
  economy & 21462 & 29 & 26 & 1.115 \\
  entertainment & 116600 & 43 & 90 & 0.478 \\
  etc. & 68919 & 40 & 36 & 1.111 \\
  IT & 104180 & 34 & 39 & 0.872 \\
  local & 42749 & 28 & 25 & 1.120 \\
  society & 65989 & 42 & 29 & 1.448 \\
  politics & 53705 & 41 & 30 & 1.367 \\
  science & 7986 & 41 & 39 & 1.051 \\
  sports & 28681 & 24 & 34 & 0.706 \\
  world & 53652 & 35 & 26 & 1.346 \\
  \hline \hline
  Total & 567384 & 37 & 37 & 1.000 \\
  \hline
  \end{tabular}
  }
 \end{table}

\subsection{Summary}

The purpose of this study is to clarify the characteristics of RTs and Likes in reference tweets.
From the basic statistics, it was confirmed that tweets containing news had characteristics that were different from those of ordinary tweets when it came to retweeting and Liking tendencies.
Moreover, the characteristics also differed greatly depending on the category.

To understand this characteristic in more detail, we compared the RTs and Likes in each category.
In this study, we aim to clarify what words are included in news that are easy to RT or Like.
The results are simple and explainable, so they are useful not only in the news domain but also in other domains, such as product marketing and content creations.

We analyzed the words that appeared in the news titles for this study.
Many tweets that contained news URLs were posted only by title. 
As a result of the preliminary research, we confirmed that the words occurring in the titles composed 80\% of the words that occurred in the tweet.

\section{Proposed Metrics: \textit{expected-dif}}

First, we define notions used in this paper.
The set of news articles is $a \in A$, the set of tweets is $t \in T$, and the set of tweets referring to news article $a$ is $t \in T_{a}$.
The number of RTs of tweet $t$ is $N_{t, RT}$, and the number of Likes of tweet $t$ is $N_{t, Like}$.
The number of RTs of news $a$ is $N_{a, RT} = \sum_{t \in T_a} N_{t, RT}$, and the number of Likes of news $a$ is $N_{a, Like} = \sum_{t \in T_a} N_{t, Like}$.
We define RT and Like as a set of actions: $action \in \{RT, Like\}$.

We classify news articles $A$ into two categories: RT-dominant and Like-dominant.
RT-dominant tweets are those for which the number of RTs is greater than the number of Likes, and  the opposite is defined as Like-dominant.
Thus, The set of RT-dominant articles is $A_{RT} = \{a \in A | N_{a, RT} > N_{a, Like}\}$, and the set of Like-dominant articles is $A_{Like} = \{a \in A | N_{a, RT} < N_{a, Like}\}$.
If $N_{a, RT}$ equals $N_{a, Like}$, an article $a$ is removed from our dataset.
Thus, we classify news articles into these two groups and compare the characteristic words of each category to visualize the differences between both news articles.

To analyze the characteristic of news articles, we define an article as a set of title words (e.g., Bag of Words).
We separate the title of the article as a word set using the Japanese Tokenizer MeCab\footnote{\url{https://taku910.github.io/mecab/} (accessed 2019--12--24)}.
In addition, we use mecab-ipadic-NEologd\footnote{\url{https://github.com/neologd/mecab-ipadic-neologd} (accessed 2019--9--30)} that corresponds to the new words obtained from the Web as a word dictionary.
The word set of article $a$ is $w \in W_a$.
A set of all words is defined as $W$.
Each article is assigned as a category $c \in C$, and the set of articles of a category $c$ is $a \in A_c$.
We define the intersection between RT/Like dominance and category $c$ as $A_{c, RT} = A_c \cap A_{RT}$ and $A_{c, Like} = A_c \cap A_{Like}$.

The simplest method for obtaining characteristic words is by counting the frequency of the words and selecting words with the highest occurrence frequencies. 
However, it was difficult to achieve effective knowledge because common words occur with high frequency.
TF--IDF is a well-known method that acts as a countermeasure against this problem.
TF--IDF is a method of evaluating characteristic words by re-counting the frequency of occurrence of specific words in the chosen documents.
However, it was difficult to use TF--IDF, as there was a large number of tweets and news titles used, and each document was short.

Therefore, we employed the $\chi$-square test process, which measures whether there is a significant difference between multiple groups.
In this paper, the proposed metrics are called the \textit{expected-dif}.
We expected these metrics to show how far the measured value was from the expected frequency and evaluate each action appropriately rather than using a simple count or the ratio of the count to the occurrence frequency of all words.
The expected-dif shows how far the word $w$ appears in the set of articles $A_{c, action}$, which are RT/Like-dominant articles with category $c$ rather than other categories articles with the same action.
The definition is as follows:

\begin{multline}
    expected \mathchar`-dif(c, w, action) = \\ as
   \frac{(T(w,A_{c, action})-E(c,w, action))^2}{E(c,w, action)} \label{eq:rank_score}.
\end{multline}
$T(w, A)$ is the number of word $w$ occurrences in an article set $A$, and $E(c, w, action)$ is the expected frequency of a word $w$ in a certain category $c$ with $action$.
$E(c, w, action)$ can be expressed as follows:
\begin{multline}
    \mathrm{E(c,w, action)} = \\
    \frac{\sum_{c'\in C}T(w, A_{c', action}) \times \sum_{w'\in W}T(w', A_{c, action})}{\sum_{c'\in C,{w'\in W}}T(w', A_{c', action})}\label{eq:expected}.
\end{multline}

We measured the influence of a word \textit{w} in each category on retweeting and Liking using this metric.

\section {Experiment and Analysis}

\subsection{Comparison between the proposed and baseline metrics} \label{count meaningwords section}

We suppose that, even if the news is in the same category, the action taken by the user changes depending on the topic.
Therefore, by seeing the characteristic words that appear in the title, we would like to confirm topics that are related to RTs or Likes.
Here, we confirm whether the feature words that are effective for confirming the topic could be extracted using the proposed metrics by comparing them with the extraction results obtained using other metrics.

For this analysis, we extracted words from 355,086 news titles obtained from reference tweets in 2017 using simple word frequency, TF--IDF, and the proposed metric \textit{expected-dif}.
We define a simple word frequency (count) and TF--IDF as baselines in this analysis.
When extracting the words, we removed the symbols and unknown words that were clearly noisy from the analysis.
This processing is based on the Japanese Tokenizer MeCab results.

The proposed metrics included a method for removing common words that made it difficult to interpret news title topics. 
Therefore, for the top 20 words in each method, we manually evaluated whether each word was effective in the interpretation as well as the effectiveness of the proposed method by the ratio of the effective words.
For this evaluation, we determined that a word that is useful for guessing the topic is a effective word.
Conversely, we determined that a word that has no meaning in a single word and is generally not used alone is not effective.
The discrimination criteria were based on the subjectivity of the first author, but the following criteria were mainly used for evaluation:

\begin{itemize}
    \item Was the word related to a category?
    \item If the word disappears from the title, is it likely that the meaning of the title would change?
    \item Is the word meaningful?
\end{itemize}

The evaluation results are shown in Table~\ref{count meaning words}.
Overall, the proposed method had a significantly higher percentage of effective words in all categories.
In the case of the proposed method, the percentage was 60\% or more in the 9 categories, and in six of those categories, both RT-dominant and Like-dominant tweets were 100\%.
The smallest percentage in the proposed method was 30\% for RT-dominant tweets in the etc. category.
In the case of count, the highest percentage category was RT-dominant at 35\%, followed by Like-dominant at 30\%.
Moreover, the etc. category RT-dominant and Like-dominant tweets and the local category of Like-dominant tweets had no meaning words. 
TF--IDF was slightly better than counting but much worse than our proposed metric.
Focusing on the content of the extracted words, in the case of counting and TF--IDF, most of the words were particle and suffix words for case inflection.
Meanwhile, in the case of the proposed method, most of the words were words that had meaning by themselves, and they had the impression of being used in news in that category.

By using the proposed metrics, we were able to successfully obtain words that promote retweeting and Liking in each category.

 \begin{table}[tb]
    \caption{\textbf{Percentage of the effective words extracted using each method:} the proposed method had a significantly higher percentage of effective words in all categories.}\label{count meaning words}
    \centering
\resizebox{\columnwidth}{!}{%
    \begin{tabular}{lrrrrrr}
      \multicolumn{5}{r}{percentage (\# of keywords)} \\
      \multirow{2}{*}{category} & \multicolumn{3}{c}{RT-dominant} & \multicolumn{3}{c}{Like-dominant} \\
      \cline{2-7}
     & count & TF--IDF & proposed & count & TF--IDF & proposed\\
    \hline \hline
        economy & 5 (1) & 20 (4) & 100 (20) & 0 (0) & 10 (2) & 100 (20) \\
        entertainment & 10 (2) & 30 (6) & 100 (20) & 20 (4) & 30 (6) & 90 (18) \\
        etc. & 0 (0) & 5 (1) & 30 (6) & 0 (0) & 0 (0) & 55 (11) \\
        IT & 20 (4) & 20 (4) & 80 (16) & 10 (2) & 10 (2) & 60 (12) \\
        local & 15 (3) & 20 (4) & 95 (19) & 0 (0) & 10 (2) & 95 (19) \\
        society & 20 (4) & 20 (4) & 100 (20) & 10 (2) & 10 (2) & 100 (20) \\
        politics & 25 (5) & 30 (6) & 100 (20) & 30 (6) & 30 (6) & 100 (20) \\
        science & 25 (5) & 20 (4) & 100 (20) & 20 (4) & 20 (4) & 100 (20) \\
        sports & 30 (6) & 30 (6) & 100 (20) & 25 (5) & 25 (5) & 100 (20) \\
        world & 35 (7) & 30 (6) & 100 (20) & 25 (5) & 25 (5) & 100 (20) \\
    \hline
    \end{tabular}%
    }
  \end{table}

\subsection{Examples of words using the proposed metrics} \label{example meaning or meaningless words section}

Table~\ref{example-entertainment} shows the top 10 words by our metric and baselines for the entertainment category as an example.
First, most of the words that were obtained by counting and TF--IDF are meaningless, and hence it is clear from the results that useful information cannot be obtained using these methods.
On the other hand, in the case of extraction using the proposed metrics, it can be seen that, for both RTs and Likes, many words related to the entertainment category appeared, and there were almost no meaningless words.
For example, the RT-dominant words among TV-related words, such as, ``主演'' (lead actor), ``ドラマ'' (drama), and ``アニメ'' (anime), appeared at the top of the characteristic words.
The Like-dominant words, among music-related words, such as, ``MV'' (music video), ``新曲'' (new single: new released) and ``ツアー'' (tour: live tour), appeared at the top of the characteristic words.

Even in news categories other than the entertainment category, there were many meaningless words in the extraction results using counting and TF--IDF.
Therefore, the proposed metrics are more useful than baselines for obtaining the features that appear in the news titles for each category.

Table~\ref{example-politics} shows the results that were obtained using the proposed metrics for the words of the news with the local category and the politics category.

In the local category, the characteristic words of the RT-dominant news were related to crimes, such as ``大阪府警'' (Osaka Prefectural Police) and ``逮捕'' (arrest), as well as words related to local sports and local tourist spots, such as ``べガルダ'' (Vegalta: soccer team name), ``J1'' (J1: soccer league name), and ``富士山'' (Mt. Fuji).

Meanwhile, in the politics category, there were many of the same words in both RT-dominant and Like-dominant tweets.
Thus, our metrics could not find the difference between RTs and Likes in this category.
This category is one of the examples that is difficult to interpret by our metrics.

From these results, it was confirmed that the characteristic words were different between RT-dominant tweets and Like-dominant tweets in the entertainment and local categories when using the proposed metrics. 
However, there was not much difference in the politics category.
More detail is provided in the next section.

\begin{table}[tb]
\centering
  \caption{\textbf{The top 10 words extracted in the news titles of the entertainment category:} the proposed metrics extracted many words related to the entertainment category appeared, and there were almost no meaningless words.}\label{example-entertainment}
  \centering
\resizebox{\columnwidth}{!}{%
  \begin{tabular}{llllll}
    \multicolumn{3}{c}{RT-dominant} & \multicolumn{3}{c}{Like-dominant} \\
    \hline
     count & TF--IDF & proposed & count & TF--IDF & proposed \\
    \hline \hline
    の & の & 主演(lead actor) & の & の & さん(a honorific suffix) \\
    が & が & 連載(serial) & に & に & MV (music video) \\
    に & に & 役(role) & が & で & 映画(film) \\
    で & で & 出演(appearance(on the stage)) & で & が & 主演(head the cast) \\
    を & を & ドラマ(drama) & を & を & 新曲(new single) \\
    は & は & アニメ(anime) & と & へ & ツアー(tour) \\
    も & も & テレビアニメ(TV anime) & は & は & ら(a pluralizing suffix) \\
    と & と & マンガ(comic) & も & た & ドラマ(drama) \\
    た & た & 放送(broadcast) & た & と & ライブ(concert) \\
    さん & さん & 描く(draw) & て & も & 氏(a honorific suffix) \\
    \hline
  \end{tabular}%
 }
 \end{table}
  
 \begin{table}[htb]
  \caption{\textbf{The top 10 words extracted using the proposed metrics:} the characteristic words were the different between the RT-dominant tweets and the Like-dominant tweets in the local category: however, there was not much difference in the politics category.}
  \centering
  \label{example-politics}
  \resizebox{\columnwidth}{!}{%
  \begin{tabular}{ll}
    \multicolumn{2}{c}{local}\\
    \hline
    RT-dominant & Like-dominant\\
    \hline \hline
    大阪府警(Osaka Prefectural Police) & 県(prefecture)\\
    関西(Kansai: one of the region in Japan) & 関西(Kansai: one of the region in Japan)\\
    大阪(Osaka) & 神戸(Kobe)\\
    逮捕(arrest) & ベガルタ(Vegalta: soccer team name)\\
    神戸(Kobe) & 弘前(Hirosaki: name of city)\\
    道内(within Hokkaido) & 県内(within the prefecture)\\
    容疑(suspicion) & 富士山(Mt.Fuji)\\
    兵庫県警(Hyougo Prefectural Police) & J1 (J1: soccer league name) \\
    辺野古(Henoko) & 盛岡(Morioka)\\
    沖縄(Okinawa) & 高校野球(highschool baseball)\\
    \hline
  \end{tabular}%
  } \\
 \centering
\resizebox{\columnwidth}{!}{%
    \begin{tabular}{ll}
    \multicolumn{2}{c}{politics}\\
    \hline
    RT-dominant & Like-dominant\\
    \hline \hline
    民進(The Democratic Party) & 自民(The Liberal Democratic Party) \\
    自民(The Liberal Democratic Party) & 安倍晋三首相(Prime Minister Shinzo Abe) \\
    氏(Honorific) & 衆院選(House of Representatives election) \\
    衆院選(House of Representatives election) & 官房長官(Chief Cabinet Secretary) \\
    首相(the Prime Minister) & 民進(The Democratic Party) \\
    代表(representative) & 氏(a honorific suffix) \\
    幹事長(Secretary-general) & 外相(the Foreign Minister) \\
    野党(Opposition Party) & 首相(the Prime Minister) \\
    官房長官(Chief Cabinet Secretary) & 幹事長(Secretary-general) \\
    衆院(House of Representatives) & 防衛相(the Minister of Defense) \\
    \hline
  \end{tabular}%
  }
 \end{table}
 
\subsection{Different tendencies in that the extracted words between RT-dominant and Like-dominant words} \label{dice section}

In this section, we analyze the difference between RT-dominant and Like-dominant word in this section detail.
In the previous section, we showed RT/Like-dominant words from some categories as examples; some categories have clear differences, and other categories do not.
Therefore, we measured how RT/Like-dominant words differ across categories quantitatively.

Next, we extracted the top 100 RT/Like-dominant words of each category and measured the similarity of these words.
Table~\ref{interpret words} shows the Dice coefficient for each category.
The Dice coefficient is a metric that indicates the similarity between two item sets, so the result show how different these words are.
The politics category has the highest Dice coefficient, which is 0.67, and the sports category comes in second at 0.66. 
In these categories, about 70\% RT/Like-dominant words are common for each category.
On the other hand, entertainment, local, etc., and science are categories that have a small Dice coefficients around 0.3.
Thus, some categories have different characteristics in RT/Like-dominant words, and other categories do not.

We examined the difference between RT/Like-dominant words in each category and interpreted the meanings of RT/Like dominant words. 
The results are shown in Table~\ref{interpret words}.
Each category has original characteristics between RT/Like-dominant words.
A discussion of these results is provided in Section~\ref{sec:discussion}.

\begin{table}[tb]
    \caption{\textbf{The characteristics of words difference:} each category has unique characteristics between RT/Like-dominant words.}
  \centering
  \resizebox{\columnwidth}{!}{%
    \begin{tabular}{l|c|l|l}
     & Dice coefficient & RT-dominant & Like-dominant \\
    \hline
    economy & 0.51 & company scandal & economy terms \\
    entertainment & 0.32 & anime, TV actor & music \\
    etc. & 0.27 & politics & animal, car, food \\
    IT & 0.51 & video game & anime \\
    local & 0.31 & crime & local \\
    society & 0.39 & disaster & crime, social event\\
    politics & 0.67 & event related government & diplomacy \\
    science & 0.33 & environment & discovery \\
    sports & 0.66 & athlete injury/retirement & game result (victory) \\
    world & 0.58 & missile (North Korea), civic movement& diplomacy\\\hline
\end{tabular}%
}
\label{interpret words}
\end{table} 

\subsection{Time series trends in characteristic words in the entertainment category} \label{time-serise}

Next, we confirmed whether characteristics could be obtained by changing the time series.
Reference tweets from January 2017 were compared weekly to analyze how that change occurred.
Table~\ref{example time} shows the analysis results.
Since week 1 was the first week of the New Year, there were many words related to the New Year TV program.
However, in terms of RT-dominant words, ``桜井'', ``桜井誠'' (a political activist), and ``韓国'' (Korea) appeared.
In that week, since ``桜井誠'' made controversial statements about the Korean musician who appeared on the TV program, these trends were affected by this event.

From week 2 onwards, the names (``狩野英孝'', ``松方弘樹'') of celebrities who had scandals and obituaries came to the top of the RT-dominant list.
In terms of Like-dominant words, music-related names, such as musician's names (``嵐'', ``宇多田ヒカル''), and movie-related words such as, ``日本アカデミー賞'' (The Japan Academy Film Prize) and ``ナラタージュ'' (Narratage: movie) appeared, while words related to scandals and gossip did not appear.

Based on this result, gossip is more Likely to be spread for the same entertainment news.
In addition, since the characteristic words differed for each week, our metrics catch the time series changes in the entertainment category.
In this way, it was possible to clarify how the characteristic words of each week were related to current events for retweeting and Liking.

\begin{table*}[tb]
  \caption{\textbf{Weekly changes in characteristic words in the entertainment category:} since the words are replaced every week, the proposed metrics catch the time series changes. Not translated words are meaningless word.}
  \centering
 \scalebox{0.85}{
  \begin{tabular}[t]{l|l|l|l}
  \multicolumn{4}{c}{RT-dominant}\\
  \hline
  week 1 & week 2 & week 3 & week 4\\
  \hline \hline
  氏(a honorific suffix) & 狩野英孝(Kano Eiko: a comedian) & 個(a counter suffix) & 個(a counter suffix) \\
  韓国(Korea) & 4月(April) & 松方弘樹(Matsukata Hiroki: an actor) & 松方弘樹(Matsukata Hiroki: an actor)\\
  歩(step) & 個(a counter suffix) & i☆Ris (an idol group) & 白熱電球(incandescent bulb) \\
  一(one) & 会見(press conference) & Shining Star (an idol group) & 交換(exchange)\\
  レコード大賞(The Japan Record Award) & 3度目(third times) & OP (opening credits) & 良い(good)\\
  頑張っ & 会長(chairman) & LED電球(LED bulb) & LED電球(LED bulb)\\
  引か & トランプ氏(Mr. Trump) & 良い(good) & 方\\
  桜井(Sakurai: a family name) & 松方弘樹(Matsukata Hiroki: an actor) & 交換(exchange) & 半数(half)\\
  最優秀新人賞(Best New Artist Award) & ヒーロー(hero) & 白熱電球(incandescent bulb) & 他人(another person)\\
  桜井誠(Sakurai Makoto: a political activist) & 日本(Japan) & 他人(another person) & ネガティブ(negative)\\
  \hline
  \end{tabular}
 }
  \label{example time}
\end{table*}

\begin{table*}[tb]
  \centering
  \begin{tabular}[t]{l|l}
  \multicolumn{2}{c}{Like-dominant} \\
  \hline
  week 1 & week 2\\
  \hline \hline
  新年(New Year) & 松坂桃李(Matsuzaka Touri: an actor)) \\
  新春(New Year) & RADWIMPS (a rock band)\\
  白(Kouhaku: music TV program) & 日本アカデミー賞(The Japan Academy Film Prize) \\
  2017年(the year 2017) & 三浦大知(Miura Daichi: a dancer)\\
  初詣(The first shrine(temple) visit of the year) & 北米(North America) \\
  嵐(Arashi: an idol group) & 妊娠(pregnancy)) \\
  ハロプロ(Hello Project: an entertainment agency) & 2017年(the year 2017) \\
  アイドル(idol) & ナラタージュ(Narratage: a movie)\\
  初日の出(the first sunrise of the year) & 新成人(new adult) \\
  音頭(Lead (in a cheer, toast, song, etc.)) & 宇多田ヒカル(Utada Hikaru: a singer-songwriter) \\
  \hline
  \end{tabular}
  \label{example time}
\end{table*}

\begin{table*}[tb]
\centering
  \begin{tabular}[t]{l|l}
  \hline
  week 3 & week 4\\
  \hline \hline
  ECD (a musician) & ECD (a musician) \\
  日本アカデミー賞(The Japan Academy Film Prize) & 三浦大知(Miura Daichi: a dancer) \\
  北米(North America) & ネガティブ(negative) \\
  他人(another person) & 良い(good) \\
  良い(good) & 他人(another person) \\
  ネガティブ(negative) & 半数(half) \\
  半数(half) & 職場(workplace) \\
  ナラタージュ(Narratage: movie) & 2017年(the year 2017) \\
  職場(workplace) & 松坂桃李(Matsuzaka Touri: an actor) \\
  怒り(anger) & ビキニ(bikini) \\
  \hline
  \end{tabular}
  \label{example time}
\end{table*}

\section{Discussion and Limitations}\label{sec:discussion}

We found the characteristic RTs and Likes in each category.
In particular, we discussed the entertainment category in Sections~\ref{count meaningwords section}, \ref{example meaning or meaningless words section}, \ref{dice section}, and \ref{time-serise}.
In the entertainment category, TV-related words appeared as RT-dominant terms, and music-related words appeared as Like-dominant terms.
This result suggests that people's sharing behavior changes according to the type of entertainment. 
In other words, people may be more aware of the type of news than the value of the news when deciding whether to RT or Like.
In the case of obtained keywords, by aggregating by time series and using proposed metrics, the resulting words indicated trend tendency (Section~\ref{time-serise}).
In addition, we confirmed that the characteristic words of each week were related to current events and retweeting and Liking.
According to these results, our metrics catch the time series changes in the entertainment category.

Although we discussed only the entertainment category due to space limitations, we also found interesting characteristics in other categories.

Table~\ref{count meaning words} in Section~\ref{count meaningwords section} shows all the cases; the extraction results contained many meaningless words in the etc. category.
Since the etc. category is a miscellaneous category, we found that there were few characteristic words unique to this category.
Thus, our proposed method picks up meaningless words as characteristic words.

The limitation of our proposed metrics is that the validation depends on the interpretability of a user who uses the metrics.
The results represent the word list to be scored by the proposed metrics, and the interpretation of the results is left up to the user.
In addition, the user requires manual work such as removing meaningless words and words that commonly appear in RT-dominant and Like-dominant tweets.
Furthermore, since our metrics require a sufficient number of tweets and articles, it is difficult to obtain results for detailed topics.
Moreover, the etc. category also fails to provide interpretable results. 
Thus, the interpretability of the results depends on how news categories are classified. 

\section{Conclusion}

The purpose of this study was to clarify the differences in retweeting and Liking.
In this paper, to analyze the difference between RTs and Likes on Twitter, we focused on word appearances in news titles.
First, we calculated basic statistics and confirmed that tweets containing news URLs have different RT and Like tendencies compared to other tweets.
It was suggested that a user might behave differently when interacting with tweets including news URLs (NewsURLs).
Next, we compared RTs and Likes for each category and confirmed that the tendency of categories is different.
Therefore, we proposed metrics for clarifying the difference in social media actions for each category used in the $\chi$-square test in order to perform an analysis focusing on the topic. 
The proposed metrics clarified the difference between words including news that is easy to RT and words including news that is easy to Like in some categories.
However there are some categories, such as politics, where the difference is difficult to understand.
Using twitter by aggregating tweets chronologically, the results indicated that it is possible to obtain keywords for grasp past trend tendencies using the proposed metrics, depending on the category.

The proposed metrics have some problems.
The results depend on how the news categories are divided, and the interpretation of the results requires manual work, so further research is necessary.

\bibliographystyle{IEEEtran}
\bibliography{wi.bib}

\end{document}